# Room-temperature intrinsic ferromagnetism of two-dimensional Na$_2$Cl crystals originated by *s*- and *p*-orbitals

**Authors:** Yimin Zhao[1†], Yunzhang Li[2†], Liang Chen[3,4†], Quan Zhang[1†], Xing Liu[2], Yizhou Yang[5], Ruobing Yi[1,4], Lihao Zhang[1], Zhenglin He[2], Yinwei Qiu[4], Dexi Nie[1], Fanghua Tian[1], Zhe Wang[1], Zhiwei Yang[1], Lei Zhang[1*], Guosheng Shi[2*], Shengli Zhang[1*] & Haiping Fang[5*]

**Affiliations:**
[1]*MOE Key Laboratory for Nonequilibrium Synthesis and Modulation of Condensed Matter, School of Physics, Xi'an Jiaotong University, Xi'an 710049, China*
[2]*Shanghai Applied Radiation Institute, Shanghai University, Shanghai 200444, China*
[3]*School of Physical Science and Technology, Ningbo University, Ningbo 315211, China*
[4]*Department of Optical Engineering, Zhejiang Prov Key Lab Carbon Cycling Forest Ecosy, Zhejiang Prov Key Lab of Chemical Utilization of Forestry Biomass, Zhejiang A&F University, Lin'an 311300, China*
[5]*School of Science, East China University of Science and Technology, Shanghai 200237, China*

\*Correspondence to: fanghaiping@sinap.ac.cn (H.F.); zhangsl@xjtu.edu.cn (S.Z.); gsshi@shu.edu.cn (G.S.); zhangleio@xjtu.edu.cn (L.Z.)

†These authors contributed equally to this work.

Ferromagnetism, as one of the most valuable properties of materials, has attracted sustained and widespread interest in basic and applied research from ancient compasses to modern electronic devices[1-7]. Traditionally, intrinsic ferromagnetism has been attributed to the permanent magnetic moment induced by partially filled *d*- or *f*-orbitals[8]. However, the development of ferromagnetic materials has been limited by this electronic structure convention. Thus, the identification of additional materials that are not constrained by this conventional rule but also exhibit intrinsic ferromagnetism is highly expected and may impact all the fields based on ferromagnetism. Here, we report the direct observation of room-temperature intrinsic ferromagnetism in two-dimensional (2D) Na$_2$Cl crystals, in which there are only partially filled *s*- and *p*-orbitals rather than *d*- or *f*-orbitals, using the superconducting quantum interference device (SQUID) and magnetic force microscope (MFM). These Na$_2$Cl crystals formed in reduced graphene oxide (rGO) membranes have an unconventional stoichiometric structure leading to unique electron and spin distributions. And the structure of these 2D Na$_2$Cl crystals, including the Na and Cl sites, is characterized *in situ* for the first time and directly observed by cryo-electron microscopy (cryo-EM) based on the observed difference in contrast between Na stacked with Cl and single Na.

**These findings break the conventional rule of intrinsic ferromagnetism and provide new insights into the design of novel magnetic and electronic devices and transistors with a size down to the atomic scale.**

Traditionally, materials with partially filled *d*- or *f*-orbitals guides the research and design[9-11] of ferromagnetic materials, but this convention seriously restricts exploration in all the related areas. Although some recent theoretical studies have mentioned magnetic properties induced by *p*-orbitals[12,13], relevant experimental results have not yet been reported, which highlights the necessity and importance of finding new materials with special electronic structures. Recently reported two-dimensional (2D) crystals[14,15] with unconventional stoichiometry and unique electron and spin distributions are highly potential candidates for new ferromagnetic materials. Here, we report the room-temperature intrinsic ferromagnetism of 2D $Na_2Cl$ crystals induced by *s*-orbitals in Na and *p*-orbitals in Cl. The structure of the 2D $Na_2Cl$ crystals formed in reduced graphene oxide (rGO) membranes is validated experimentally for the first time by optimized cryo-electron microscopy (cryo-EM), which confirms the atomic configuration of the $Na_2Cl$ crystals based on the observed difference in contrast between Na stacked with Cl and single Na. Theoretical computations suggest that the origin of the magnetism is the unconventional stoichiometric structure of $Na_2Cl$. In contrast to that of the structure of ordinary NaCl, the total density of states (DOS) of these $Na_2Cl$ crystals, including the spin-up and spin-down DOS, shows high asymmetry.

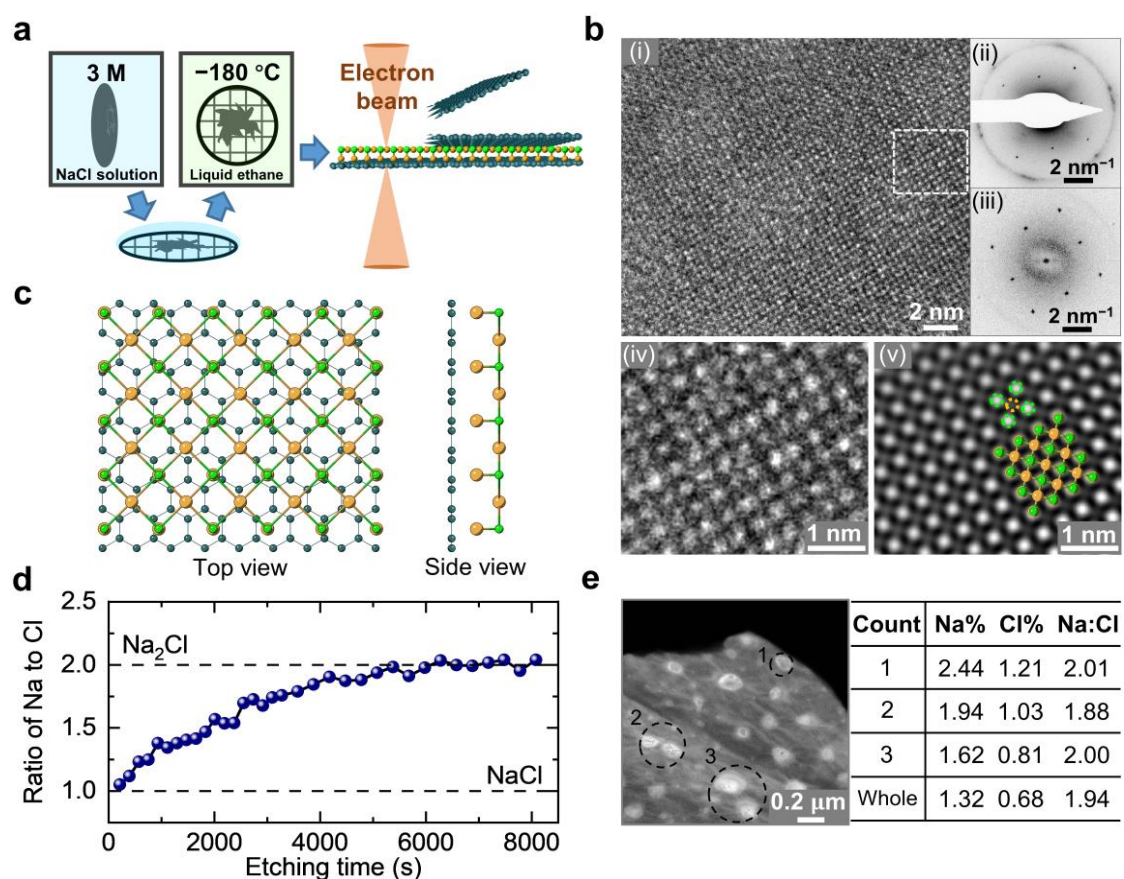

**Fig. 1 | 2D Na–Cl crystals formed in rGO membranes. a,** Schematic diagram of the sample preparation process and cryo-EM characterization. **b,** (i) Cryo-EM images of Na–Cl crystals

observed in the rGO membranes. (ii) Diffraction pattern of the crystal structure obtained in the region marked in (i) in electron diffraction mode. (iii) Fast Fourier transform (FFT) of the region marked in (i), showing the same square lattice observed in (ii). (iv) Magnified image of the boxed region in (i). (v) Background-removed inverse Fourier transform image of the image shown in (iv) overlaid with a stable $Na_2Cl$ model, in which the Na and Cl atoms are shown as larger orange spheres and smaller green spheres, respectively. The bright dots surrounded by the green dashed lines correspond to the mass-thickness contrast of overlapping Na and Cl atoms, while the darker gray dot surrounded by the yellow dashed line in the center of the four bright dots corresponds to Na alone. The distance between the bright dots and the nearest-neighbor gray dots is 2.81±0.03 Å, consistent with the bond length of the ordinary NaCl (200) crystal surface (2.81 Å). **c,** One stable structure of $Na_2Cl$ crystal modules adsorbed on a graphene sheet[14]. Na, Cl and C atoms are displayed in orange, green and dark blue, respectively. **d,** Atomic ratio of Na to Cl as a function of the etching time obtained by the argon ion etching of dried Na–Cl@rGO membranes by the XPS experiment. The lower and upper dashed lines represent the atomic ratios in NaCl and $Na_2Cl$, respectively. **e,** High-angle annular dark-field scanning transmission electron microscopy (HAADF-STEM) image obtained at the edge of the dried Na–Cl@rGO membranes. The table on the right shows the atomic percentages of Na and Cl in the selected areas of the dried Na–Cl@rGO membranes.

Freestanding rGO membranes[14,16,17] were prepared from an rGO suspension via the drop-casting method as described in ref[14]. These membranes were incubated in a 3.0 mol/L (M) NaCl solution overnight under ambient conditions, and then split into small pieces to expose the Na–Cl crystals formed between the membrane layers. The membranes coated with original salt solution were plunge-frozen in liquid ethane at 20°C under 100% humidity and analyzed by cryo-EM[18-20] to study the *in situ* formation[21,22] of Na–Cl crystals (Fig. 1a). As the Na–Cl crystals are 2D and very sensitive to high-energy electrons, the cryo-EM measurement conditions must be optimized. We used low-dose mode and an extremely short exposure time (~0.2 s) at a 200 kV TEM high tension, which showed the best balance between the electron beam damage and the resolution among different exposure times and high tensions. As a result, at the edge of the rGO membranes, relatively thin (few layers of rGO sheets) membranes with Na–Cl crystal domains that are suitable for TEM imging could be found (Supplementary Fig. 1).

Within several such regions, stable single-crystal diffraction patterns of the Na–Cl crystals, other than the diffraction patterns of graphene, could be directly observed (Fig. 1b). Electron diffraction and FFT analyses[23] of the observed Na–Cl lattice show that these crystals have a square structure with a lattice spacing of 3.97±0.03 Å, consistent with the single-layer NaCl (200) crystal surface [Fig. 1b (ii, iii)]. To further investigate the bond form and element composition inside the crystal, we look for clues from high-resolution atomic images. However, since the samples are very sensitive to electron irradiation, these images are susceptible to noise interference even in low-dose mode and under an extremely short exposure time; thus, these original images were processed by the background-removed inverse Fourier transform method. As a result, a clearly darker gray dot surrounded by four bright dots is observed [Fig. 1b (v)], and the distance from each gray dot to the nearest bright dot is 2.81±0.03 Å, consistent with the bond length of the ordinary NaCl (200) crystal surface (2.81 Å). Importantly, we note that the difference in contrast between the bright dots and gray dots is consistent with the simulated TEM image[24] of the previously predicted 2D $Na_2Cl$ model[14] (Fig. 1c), and is fundamentally different from the atomic image of the ordinary NaCl (200) crystal surface obtained with the same experimental parameters (Supplementary Fig. 2). In addition, the experimental cryo-EM images of Na–Cl crystals are also significantly different from the simulated TEM images of ordinary single-layer or bulk NaCl crystals (Supplementary Fig. 3), suggesting that the observed crystals may be $Na_2Cl$ rather than ordinary NaCl crystals. Notably, atomic-resolution images for such radiation-sensitive 2D crystals were successfully obtained by performing cryo-EM without a drying step to avoid ordinary bulk NaCl crystals, taking advantage of the significant mass difference between a single Na atom and overlapping Na and Cl atoms, and applying optimized cryo-EM imaging conditions.

To further confirm that the obtained crystals are $Na_2Cl$, X-ray photoelectron spectroscopy (XPS) was used to measure the atomic ratio of Na to Cl in dried rGO membranes containing Na and Cl, namely, *the dried Na–Cl@rGO membranes*. The dried Na–Cl@rGO membranes were obtained by first incubating rGO membranes in 3.0 M NaCl solutions for 24 hours, centrifuging them to remove the free solution and drying them at 70°C for 2 hours. We note that the same Na–Cl crystals shown in Fig. 1b were also identified in the dried Na–Cl@rGO membranes (Supplementary Fig. 2). By argon ion etching, the depth profile obtained by XPS experiment shows that the ratio of Na to Cl changes from ~1:1 in the surface of the rGO membranes to a stable ~2:1 ratio in the inner sheets (Fig. 1d). The membrane surface may have ordinary NaCl due to evaporation of the adsorbed salt solution, but in the inner sheets, mainly Na–Cl crystals with a stable Na:Cl ratio of ~2:1 are present. The existence of $Na_2Cl$ was also confirmed by conventional (noncryogenic) TEM energy-dispersive X-ray spectroscopy (EDS) analysis. As shown in the HAADF-STEM image in Fig. 1e, the crystal domain was mainly composed of Na and Cl with an atomic ratio of ~2:1. In addition, the EDS results obtained for multiple areas of the thin edge also showed that the atomic ratio of Na:Cl varied between 1 and 2 (Supplementary Fig. 4), consistent with the experimental results obtained by XPS. And X-ray diffraction (XRD) experiments on the dried Na–Cl@rGO membranes further confirmed that the crystals formed in the rGO membranes are 2D $Na_2Cl$ crystals corresponding to the NaCl (200) crystal surface (Supplementary Fig. 5).

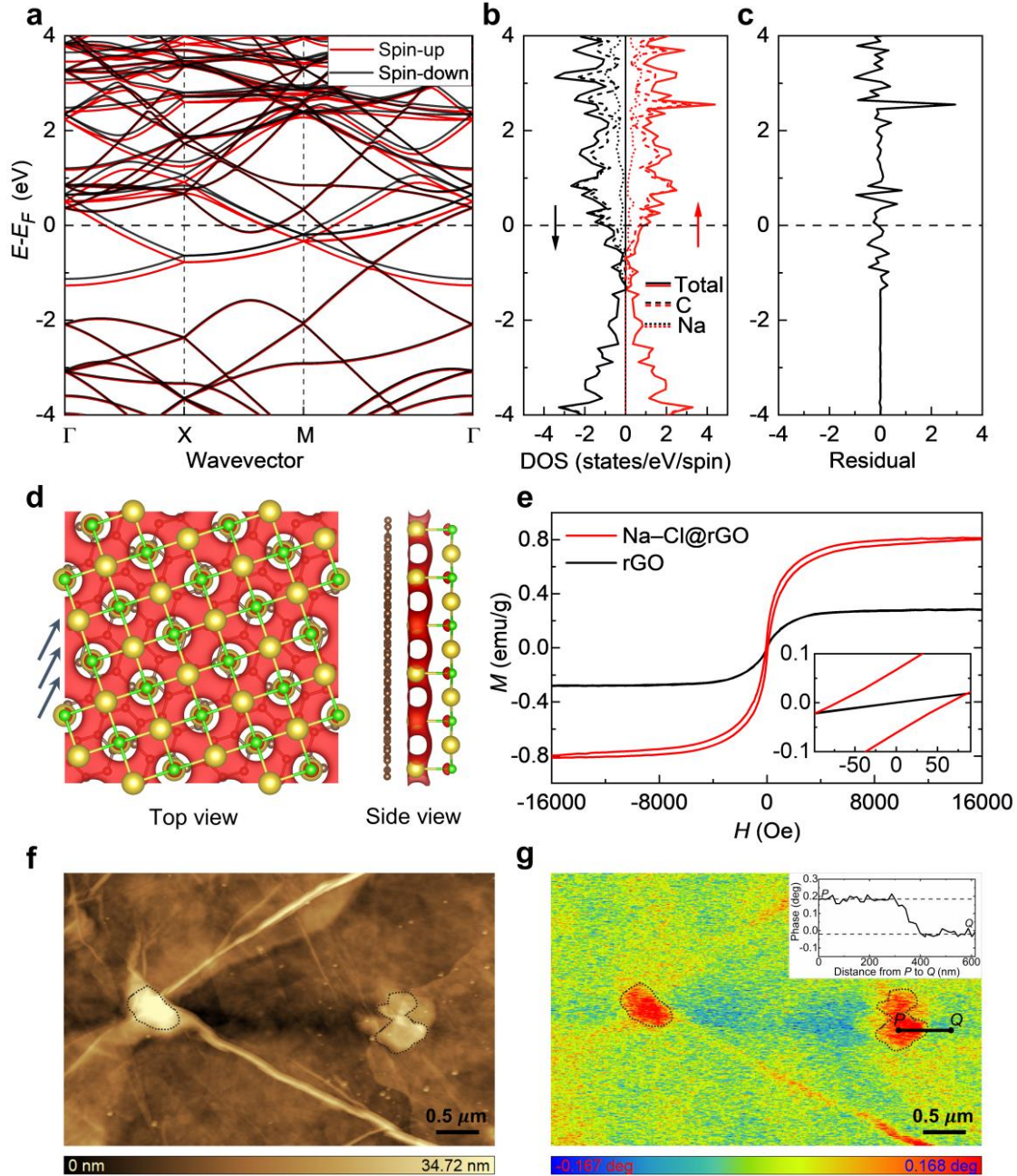

**Fig. 2 | Magnetic analysis of 2D Na$_2$Cl crystals.** Spin-resolved band structures (**a**), spin-polarized DOS of graphene-Na$_2$Cl (**b**), and difference in spin-polarized DOS (**c**). **d**, Spin charge distribution of the graphene-Na$_2$Cl structure. For ease of display, the side view is taken along the direction of the arrows in the top view. **e**, Hysteresis loops of dried Na–Cl@rGO membranes and dried rGO membranes measured at 300 K. The experimental magnetic field is perpendicular to the sample surface. **f**, AFM topography image of Na–Cl@rGO membranes obtained by the mechanical peeling method. **g**, MFM image obtained in the same area as the AFM image after raising the scanning probe by 70 nm. Inset: Phase changes with the distance from $P$ (a point in the crystal domain) to $Q$ (a point in the pure rGO area). The two Na$_2$Cl crystal domains in **f** and **g** circled by black dashed lines mark the positions where the phase decreases by half from the crystal domain to the pure rGO area.

To study the magnetic properties of the prepared 2D $Na_2Cl$ crystals, we carried out a magnetic-related theoretical analysis of $Na_2Cl$ crystals by spin-polarized density functional theory (DFT) calculations[25-27]. We note that there has been a report[28] that ordinary NaCl has a nonmagnetic ground state ($M_{tot}$ = 0) and that the total DOS based on calculations is perfectly symmetric, indicating that a system consisting of ordinary NaCl crystals is nonmagnetic. However, here, for the graphene-$Na_2Cl$ system, the spin-resolved band structure (Fig. 2a) shows that the energy bands of the spin-up and spin-down electrons do not coincide and that the total DOS of the $Na_2Cl$ crystal, including the spin-up and spin-down DOS, is considerably asymmetric (Fig. 2b-c), indicating that there is intrinsic magnetism in the graphene-$Na_2Cl$ system (~0.053 μB per Na atom). Fig. 2d shows that the spin density is mainly concentrated in the vicinity of Na with dangling bonds and adjacent Cl, so the $Na_2Cl$ crystals introduce new spin centers into the Na–Cl@rGO membranes and contribute significant net moments, indicating that the magnetism is primarily induced by *s*-orbitals in Na and *p*-orbitals in Cl.

To experimentally confirm this intrinsic magnetism, magnetic characterization by the superconducting quantum interference device (SQUID) was performed. The hysteresis loops of the dried Na–Cl@rGO membranes and dried rGO membranes at 300 K (Fig. 2e) show that the saturation magnetic moment ($M_s$) of the dried Na–Cl@rGO membranes ($M_s$ = ~0.81 emu/g) is approximately 3 times that of the dried rGO membranes ($M_s$ = ~0.28 emu/g) under the same experimental conditions. Considering that the mass percentage of sodium is only ~4.3%, the presence of $Na_2Cl$ crystals greatly enhances the ferromagnetism of the membrane (~0.051 μB per Na atom), consistent with the theoretical results (~0.053 μB per Na atom) (details in Methods), indicating the room-temperature ferromagnetism of 2D $Na_2Cl$ crystals.

To exclude the possibility that this ferromagnetism is caused by the edge or defect effects[29-31] of the $Na_2Cl$ crystals, the ferromagnetic signals of the crystal domains inside the Na–Cl@rGO membranes were also detected with magnetic force microscopy (MFM) (Fig. 2f-g). The dried Na–Cl@rGO membranes, which were the same as those used for the SQUID measurements, were first uniformly peeled by the mechanical peeling method[32] using adhesive tape to expose the crystal domains inside the membranes (Supplementary Fig. 6) and then quickly transferred to a clean silicon wafer to provide a flat substrate for MFM probe scanning (Supplementary Fig. 7). The atomic force microscopy (AFM) image in Fig. 2f shows a topography image of the rGO flake and Na–Cl crystal domains on these flat rGO membranes. The MFM image in Fig. 2g was obtained in the same area as the AFM image, with a lift scan height of 70 nm, where the long-distance van der Waals force is negligible and the magnetic force is dominant. Weak magnetic signals were observed in the flake domains, consistent with a previous study of the magnetic properties of graphene[33]. Importantly, strong magnetic signals clearly fill the $Na_2Cl$ crystal domains without obvious enhancement on the edge or at certain spots of the domain, indicating that the magnetism of $Na_2Cl$ is intrinsic instead of due to edge or defect effects (Fig. 2g, Supplementary Fig. 7). Overall, the experimental results from SQUID and MFM together demonstrate the existence of room-temperature intrinsic ferromagnetism in 2D $Na_2Cl$ crystals predicted by theoretical calculations.

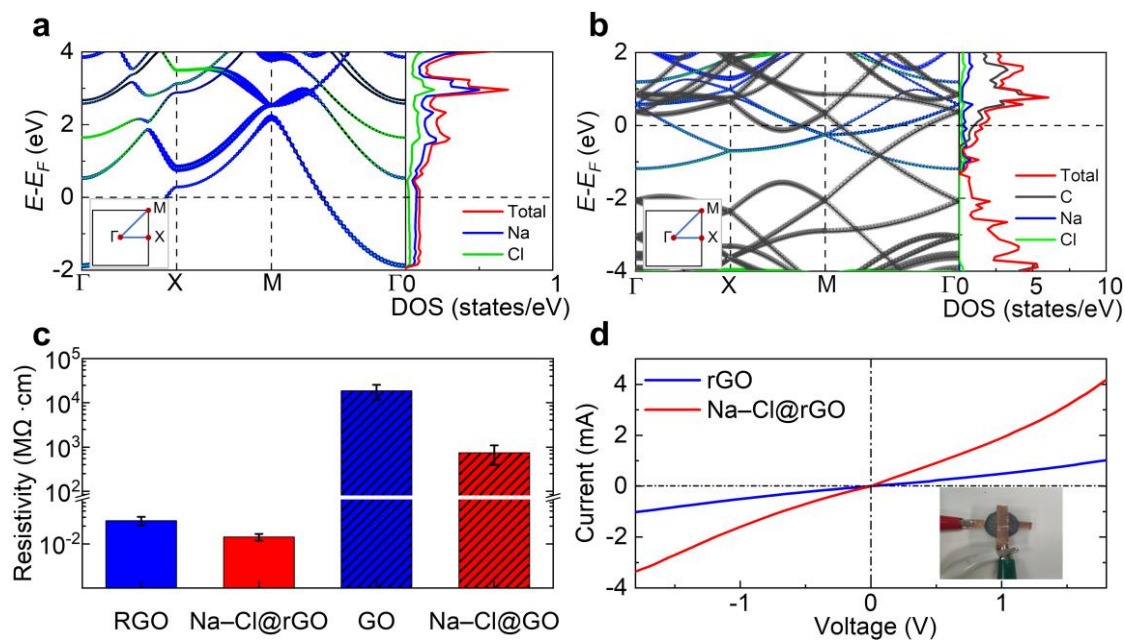

**Fig. 3 | Metallicity analysis and heterojunction behavior of 2D Na₂Cl crystals. a,** Projected band structure and projected DOS on each element in the primitive cell of Na₂Cl alone. **b,** Projected band structure and projected DOS on each element in the primitive cell of the full model, as shown in Fig. 1c. The weight of the contribution of each element in (a) and (b) is represented by the size of the points. The inset shows the high symmetry q-point path in the Brillouin zone, where $E_F$ refers to the Fermi level. **c,** Resistivity of the samples obtained with a multimeter (Supplementary Fig. 8). Dried rGO membranes, GO membranes, Na–Cl@rGO membranes and Na–Cl@GO membranes were selected as experimental objects, and two electrodes were connected to their upper and lower surfaces. **d,** Current-voltage curves obtained on Na–Cl@rGO membranes (red) and rGO membranes (blue). Inset: Photograph of the experimental setup for measuring current-voltage curves with copper foil electrodes connected to dried membranes.

Due to the unique electron and spin distributions of the Na₂Cl crystals, these materials also show novel electrical properties, such as metallicity and heterojunction behavior. The metallicity of Na₂Cl crystals was first investigated by theoretical calculations. The projected electron band structure and DOS of Na₂Cl crystals show that these crystals have distinct metallicity and that Na plays a major role in this property (Fig. 3a). The complexes of graphene with Na₂Cl crystals also have strong metallicity, and Na and Cl together contribute ~35.4% of the total electronic states near the Fermi level, indicating that the Na₂Cl crystals are also metallic in this system (Fig. 3b). To experimentally confirm the metallicity of the Na₂Cl crystals, the resistivities of the dried rGO membranes and Na–Cl@rGO membranes were measured with a multimeter. The results show that the resistivities of these samples are all less than $4.0\times10^{-2}$ MΩ·cm, indicating that they are conductive (Fig. 3c) and we note that the metallicity of rGO membranes prepared by the same method has been reported in the literature[34]. In contrast to that of dried GO membranes with high resistivity, the average resistivity of the dried Na–Cl@GO membranes was reduced by at least one order of magnitude, while the resistivity of the dried Na–Cl@rGO membranes was more than 2 times less than that of pure dried rGO membranes (Fig. 3c, Supplementary Table 1). The heterojunction behavior of graphene–$Na_2Cl$ was also experimentally demonstrated. As shown in Fig. 3d, compared with that of the dried rGO membranes, the current-voltage curve of the dried Na–Cl@rGO membranes under positive and negative voltages exhibits obvious asymmetric behavior, presenting a typical rectification characteristic.

To summarize, for 2D $Na_2Cl$ crystals, in which there are no partially filled *d*- or *f*-orbitals, DFT computations show the existence of an asymmetric spin-polarized DOS for the *s*-orbitals in Na and *p*-orbitals in Cl, which induce intrinsic ferromagnetism. SQUID measurements demonstrate the room-temperature ferromagnetism of the 2D $Na_2Cl$ crystals, and MFM measurements show that the strong magnetic signals clearly fill the entire $Na_2Cl$ crystal domains, suggesting that the magnetism is intrinsic rather than due to the effects of edges or defects. Compared to heavy metal magnetic materials, such room-temperature intrinsic ferromagnetic material induced by *s*- and *p*-orbitals breaks through the traditional rule that intrinsic ferromagnetism can only be induced by *d*- or *f*-orbitals, and extends the magnetic research to more elements (such as the main group elements), which offers new inspiration and insight to the development of ferromagnetism[35-37] and stimulates the discovery of more new intrinsic ferromagnetic states and corresponding novel electromagnetic applications[7,38-41]. Furthermore, this intrinsic ferromagnetic material prepared from a simple salt is economical and environmentally and ecologically friendly for use on a much larger scale. This finding advances the understanding of magnetic mechanisms in nature, providing insight into the origin of natural magnetism of compounds in simple salt solution systems under ambient conditions.

Although 2D Na–Cl and K–Cl systems with abnormal stoichiometries have been recently reported[14], the discovery was based only on macroscopic characterization, and the crystal configuration as well as the resulting novel properties have not been further experimentally revealed. Here, by optimized cryo-EM, based on the observed difference in contrast between Na stacked with Cl and single Na, the atomic-resolution structure of the prepared 2D $Na_2Cl$ crystals, including the Na and Cl sites, were characterized *in situ* and directly observed in rGO membranes in unsaturated NaCl solutions under ambient conditions. Clearly, the configuration of these 2D $Na_2Cl$ crystals is significantly different from that of ordinary bulk or single-layer NaCl crystals. The experimental observations of other novel properties, including metallicity and heterojunction behavior, further demonstrate the existence of a unique electron distribution and the extension of applications of 2D $Na_2Cl$ crystals in rGO membranes, which could greatly facilitate the design of graphene-based spintronic devices and nanoscale transistors as well as sensors.

## Methods

**Fabrication of freestanding rGO membranes.** The reduced graphene oxides(rGO) membranes were prepared from natural graphite powders via a modified Hummers method as described previously[14,16,17]. The prepared rGO membranes were stored in dried and clean vacuum containers before usage.

**Fabrication of cryo-EM and dried specimens.** Cryo-EM specimens were prepared by immersing rGO membranes in 3.0 M (mol/L) NaCl solution. In brief, the freestanding rGO membranes were immersed in NaCl solution of 3.0 M concentration overnight at room temperature. After that, the wet membranes were manually torn into small pieces in the same solution. One piece of the rGO membrane containing the salt solution was then moved and fixed to a membrane-free Au TEM Grid. By using a FEI Vitrobot rapid-plunging device with conditions set at 100% humidity and room temperature, the sample was blotted with filter paper (#595, Schleicher & Schuell) for 2 s and then flash-frozen in liquid ethane. The frozen-hydrated specimen was then transferred to liquid nitrogen and placed onto a Gatan 626 cryo-TEM holder while immersed in liquid nitrogen and isolated from the environment by a tightly closed shutter. During insertion into the TEM column, temperatures did not increase above -170 °C.

Dried membranes with Na and Cl for conventional (non-cryogenic) TEM, XPS, EDS, room-temperature ferromagnetism, electric resistivity and heterojunction behavior measurement were prepared by first incubating the rGO membranes in 3.0 M NaCl solutions overnight and then centrifuging to remove the free solution and drying at 70 °C for 2 hours in vacuum drying oven.

**TEM data collection.** Cryo-EM micrographs were acquired at −180 °C on a FEI CETA 4k × 4k CMOS camera by a FEI F200C transmission electron microscope (TEM) operating at 200-kV and low-electron-dose conditions. Selected-area electron diffraction (SAED) images were taken with a ~350 nm diameter selected-area (SA) aperture. The exposure time was varied between 0.2 and 1 s.

High-resolution atomic images were obtained under freezing conditions. When the spherical aberration ($Cs$) value and the defocus value are suitable, the bright contrast area usually corresponds to the atom. This has been confirmed in the study of atomic structure of sensitive battery materials by cryo-electron microscopy[42-44].

Conventional (non-cryogenic) high-resolution TEM micrographs and SAED images were acquired at room temperature by the same FEI F200C TEM operating at 200-kV. High-angle annular dark field scanning TEM (HADDF-STEM) and energy-dispersive X-ray spectroscopy (EDS) were performed at room temperature on a JEOL 4k × 4k CMOS camera by a JEOL JEM-F200 (HR) TEM operating at 200-kV.

**TEM image simulation by QSTEM.** To further confirm the experimental results obtained by cryo-EM, we also performed TEM simulations on the recently predicted Na$_2$Cl crystal structure[14] and the single-layer and double-deck structures of ordinary NaCl (200) crystal surface by QSTEM[24]. QSTEM is a script software that uses the multislice algorithm to perform TEM-related simulations. we performed image simulations with varying focus values and crystal thickness, assuming an accelerating voltage of 200 kV, *Cs* value of -15 μm, beam convergence angle of 0.2 mrad, and defocus of -6 nm. Other aberrations are not considered (set to defaults). Image simulations for both the atomic resolution images are presented in Supplementary Fig. 3.

The area of brightness contrast in the Extended Data Fig. 3 represents the area where atoms exist. It can be seen from the figure that for the double-layer structure of ordinary NaCl, the brightness contrast of all atomic positions is basically same. For the single-layer structure of ordinary NaCl, the brightness at the position of Cl is about 1.6 times that of Na, and the brightness difference is not obvious. But for the recently predicted two-dimensional crystal Na$_2$Cl structure, the brightness of the position where Na and Cl coexist is about 3.1 times that of the position where Na is alone. At this time, it can be seen from the image that the brightness of Na alone is not obvious. Taking into account the influence of noise and other factors in the actual electron microscope imaging, it is difficult to observe the position of Na alone.

**XPS data collection.** The dried Na–Cl@rGO membranes were characterized by X-ray photoelectron spectroscopy (XPS), which was performed on a Thermo Fisher ESCALAB Xi$^+$ system. The atomic percentage of each element in different depth of dried membranes was obtained by XPS analysis with argon ion etching. The instrument conditions were as follows: source gun type, Al K Alpha; ion energy, 3000 eV; current, 2.73 μA; raster size, 2.5 mm; spot size, 650 μm.

**XRD data collection.** To further confirm the bulk structure of the Na$_2$Cl crystals, we performed the XRD detection experiences on the dried Na–Cl@rGO membranes and pure dried rGO membranes by X-ray diffractometer *Bruker D8 ADVANCE*. After the Na–Cl@rGO membranes and pure rGO membranes were dried in a 70 °C oven under vacuum for 2 hours, they were quickly taken out and transferred to the sample stage of the X-ray diffractometer for observation.

**SEM data collection.** The dried Na–Cl@rGO membranes were also characterized by Field Emission SEM *GeminiSEM 500*. The dried Na–Cl@rGO membranes were peeled off with adhesive tape by mechanical peeling method so that the special Na–Cl crystals could be exposed, and then the peeled sample was quickly transferred to the vacuum chamber of *GeminiSEM 500* for scanning electron microscopy analysis.

**Room-temperature ferromagnetism measurement.** The magnetic properties of the dried Na–Cl@rGO membranes and pure dried rGO membranes with respect to temperature and field were measured using a quantum design MPMS-SQUID VSM-094 magnetometer (sensitivity ~1×10$^{-8}$ emu). Before membrane samples were tested, the membranes were cut into rectangular pieces (about 3 mm × 3 mm), which were weighed by using a microbalance. Then the samples were loaded into the quartz holder (provided by Quantum Design) and wrapped, after which the quartz holder was attached to the sample rod and inserted into the magnetometer. Magnetic hysteresis curves of the pure dried rGO membranes and dried Na–Cl@rGO membranes were measured at room-temperature (300 K) in the field range of −30 kOe < H < +30 kOe.

The hysteresis loops of the dried Na–Cl@rGO membranes and the dried rGO membranes at 300 K (Fig. 2e) show that the saturation magnetic moment ($M_s$) of the dried Na–Cl@rGO membranes ($M_s$ = ~0.81 emu/g) is about 3 times that of the dried rGO membranes ($M_s$ = ~0.28 emu/g) under the same experimental parameters. Considering that the mass of the dried Na–Cl@rGO membrane used for the M-H curve measurements is 0.044 mg, the mass of the pure rGO membrane is 0.043 mg, and the mass percentage of Na elements inside the Na–Cl@rGO membranes obtained by the XPS experiment is 4.31%, it can be calculated that the average magnetic moment of each Na atom in the Na$_2$Cl crystal is ~0.051 μB, consistent with the intrinsic magnetic moment of Na$_2$Cl by DFT theoretical calculation (~0.053 μB per Na atom).

To further achieve direct characterization of the magnetic properties of the Na$_2$Cl crystals inside the dried Na–Cl@rGO membranes by MFM, we uniformly peeled the dried Na–Cl@rGO membranes by mechanical peeling method with adhesive tape to expose the morphology and crystal domains inside the membranes. The torn Na–Cl@rGO membranes were then quickly transferred to the vacuum chamber of SEM for morphological characterization and EDS analysis. Experimental results show that the crystal domains distributed inside the torn dried Na–Cl@rGO membranes are mainly composed of Na and Cl, and the ratio of Na and Cl varies from 1.2 to 2 (Supplementary Fig. 6). Especially, in the thinner crystal domains (5 in Supplementary Fig. 6a and 12 in Supplementary Fig. 6b), the atomic ratio of Na to Cl is mainly ~2:1.

Since the flatness of the dried Na–Cl@rGO membranes peeled off with tape cannot meet the requirements of MFM probe scanning (Supplementary Fig. 7a), we quickly attached the exposed membrane surface to the clean silicon wafer after tearing off the dried Na–Cl@rGO membranes with tape. After removing the tape, a flat Na–Cl@rGO membrane transferred to the silicon wafer was obtained (Supplementary Fig. 7b-c). Then we performed MFM-based magnetic characterization of Na–Cl crystals distributed on the dried Na–Cl@rGO membranes. The MFM image in Supplementary Fig. 7e was obtained at the same area as the AFM image (Supplementary Fig. 7d), with a lift scan height of 50 nm, where the long-distance van der Waals force is negligible and the magnetic force is dominant, strong magnetic signals were clearly observed in the thinner crystal domains, revealing the magnetism of $Na_2Cl$ crystals.

**Electrical resistivity measurement.** Three dried GO, rGO, Na–Cl@GO, and Na–Cl@rGO membranes with the thickness of ~0.1 mm and diameter of ~15 mm were prepared, respectively. Then the electric resistivities of five selected positions of each dried membrane were measured by using the multimeter with two electrodes (~24 $mm^2$) connecting with the up and down surfaces of the membranes under ambient conditions. During the measurement, each position was exerted the same pressure in the perpendicular direction of the membranes. All the measurements were performed in a glove box filled with highly pure Ar gas ($O_2$ and $H_2O$ concentration were less than 0.1 ppm).

**Heterojunction behavior measurement.** The rectification behavior of the dried Na–Cl@rGO and pure dried rGO membranes was measured with a SourceMeter system (part number 2636B, Keithley). Corresponding current–voltage was measured at a bias voltage of 5V applied to two Cu electrodes (size of ~16 $mm^2$) connecting with the up and down surface of the membranes under ambient conditions.

**Acknowledgments:** We thank Profs. Chungang Duan and Yaohua Jiang for constructive suggestions. We also thank the Instrument Analysis Center of Xi'an Jiaotong University for cryo-EM imaging, XPS analyses, and magnetic measurement.

**Funding:** the National Natural Science Foundation of China (Nos. 11922410,11774279, 12074341 and U1932123), the Key Research Program of Frontier Sciences of the Chinese Academy of Sciences (No. QYZDJ-SSW-SLH053), the Key Research Program of the Chinese Academy of Sciences (No. KJZD-EW-M03), Shanghai Supercomputer Center of China and the Special Program for Applied Research on Super Computation of the NSFC-Guangdong Joint Fund (the second phase), the World-Class Universities (Disciplines) and the Characteristic Development Guidance Funds for the Central Universities, and Young Talent Support Plan of Xi'an Jiaotong University.

**Author contributions:** H.F., S.Z. L.Z. and G.S. designed the project. Y.Z., R.Y. and L.Z. performed the cryo-EM and conventional TEM experiments; Y.Z., L.C., Q.Z., Y.Y., Z.Y., F.T., L.Z., G.S., S.Z. and H.F. performed experiments and calculations for room-temperature ferromagnetism and MFM measurements; Y.L. L.C., Q.Z., R.Y., Z.H., X.L., S.Z. and G.S. performed experiments for electric properties. L.Z., G.S., H.F., S.Z., Y.Z., L.C., Y.L. and Q.Z. co-wrote the manuscript. All authors discussed the results and commented on the manuscript.

**Competing interests:** Authors declare no competing interests.

**Data and materials availability:** All data is available in the main text or the supplementary materials.

# Supplementary Information

## Room-temperature intrinsic ferromagnetism of two-dimensional $Na_2Cl$ crystals originated by *s*- and *p*-orbitals

**Authors:** Yimin Zhao[1†], Yunzhang Li[2†], Liang Chen[3,4†], Quan Zhang[1†], Xing Liu[2], Yizhou Yang[5], Ruobing Yi[1,4], Lihao Zhang[1], Zhenglin He[2], Yinwei Qiu[4], Dexi Nie[1], Fanghua Tian[1], Zhe Wang[1], Zhiwei Yang[1], Lei Zhang[1*], Guosheng Shi[2*], Shengli Zhang[1*] & Haiping Fang[5*]

**Affiliations:**
[1]*MOE Key Laboratory for Nonequilibrium Synthesis and Modulation of Condensed Matter, School of Physics, Xi'an Jiaotong University, Xi'an 710049, China*
[2]*Shanghai Applied Radiation Institute, Shanghai University, Shanghai 200444, China*
[3]*School of Physical Science and Technology, Ningbo University, Ningbo 315211, China*
[4]*Department of Optical Engineering, Zhejiang Prov Key Lab Carbon Cycling Forest Ecosy, Zhejiang Prov Key Lab of Chemical Utilization of Forestry Biomass, Zhejiang A&F University, Lin'an 311300, China*
[5]*School of Science, East China University of Science and Technology, Shanghai 200237, China*

*Correspondence to: fanghaiping@sinap.ac.cn (H.F.); zhangsl@xjtu.edu.cn (S.Z.); gsshi@shu.edu.cn (G.S.); zhangleio@xjtu.edu.cn (L.Z.)

†These authors contributed equally to this work.

**Theoretical computation methods:**

**Crystal structure optimization, electronic and magnetic property calculations**

First principles calculations based on the spin-polarized DFT were performed with the Vienna ab initio Simulation Package (VASP) code[1]. The Projector Augmented Wave (PAW) pseudopotentials were applied[2,3]. In order to investigate the structural properties, the generalized gradient approximation (GGA) of the Perdew–Burke–Ernzerhof (PBE) was used to treat the exchange-correlation interaction between electrons[4]. A plane-wave basis set with a kinetic-energy cut-off of 520 eV was used to expand the wave function of valence electrons ($2s^2p^2$ for C, $3s^2p^5$ for Cl and $3s^1$ for Na). Three-dimensional periodic boundary conditions were applied to simulate the infinite systems. A 20 Å vacuum space between sheets was set to prevent the interaction between two layers. The structural relaxations were performed by computing the Hellmann-Feynman forces within total energy and force convergences of $10^{-5}$ eV and $10^{-3}$ eV/Å, respectively. Gamma-centered Monkhorst-Pack grid of $5 \times 5 \times 1$ was used in the relaxed, electronic and magnetic properties calculations. In band structures calculations, the Gaussian smearing with broadening of 0.02 eV was used for partial occupancy, while the density of states was calculated by the tetrahedron method with Blöchl corrections.

**Thermodynamic stability analysis of the graphene-Na$_2$Cl crystal structures**

The ab initio molecular dynamics (AIMD) simulation was performed using VASP code[1], starting from the relaxed structure obtained by DFT calculations, with no initial velocities assigned. The size of supercell used in simulation is $17.79 \times 17.79$ Å$^2$ for model. A 20 Å vacuum space between sheets is set to prevent the interaction between two layers. Brillouin zone is sampled using only the Γ k-point. The simulations were carried out in a canonical ensemble (NVT) with a Nosé thermostat for temperature control[5]. The temperature was constant at 300 K by lasting for 10 ps with time step of 1 fs.

**Bader charge analysis**

Population analysis was performed using the Bader Charge Analysis code developed by Henkelman's group[6-8]. Most charge analysis methods such as Mulliken population analysis were based on electron wave functions and thus sensitive to the type and cutoff of basis sets. Bader analysis[9], on the other hand, makes use of zero flux surfaces and distinguish the ownership of electrons. The charge enclosed within the Bader volume is a good approximation to the total electronic charge of an atom and the analysis was merely dependent on charge density distribution. In this work, charge density distribution was calculated from VASP.

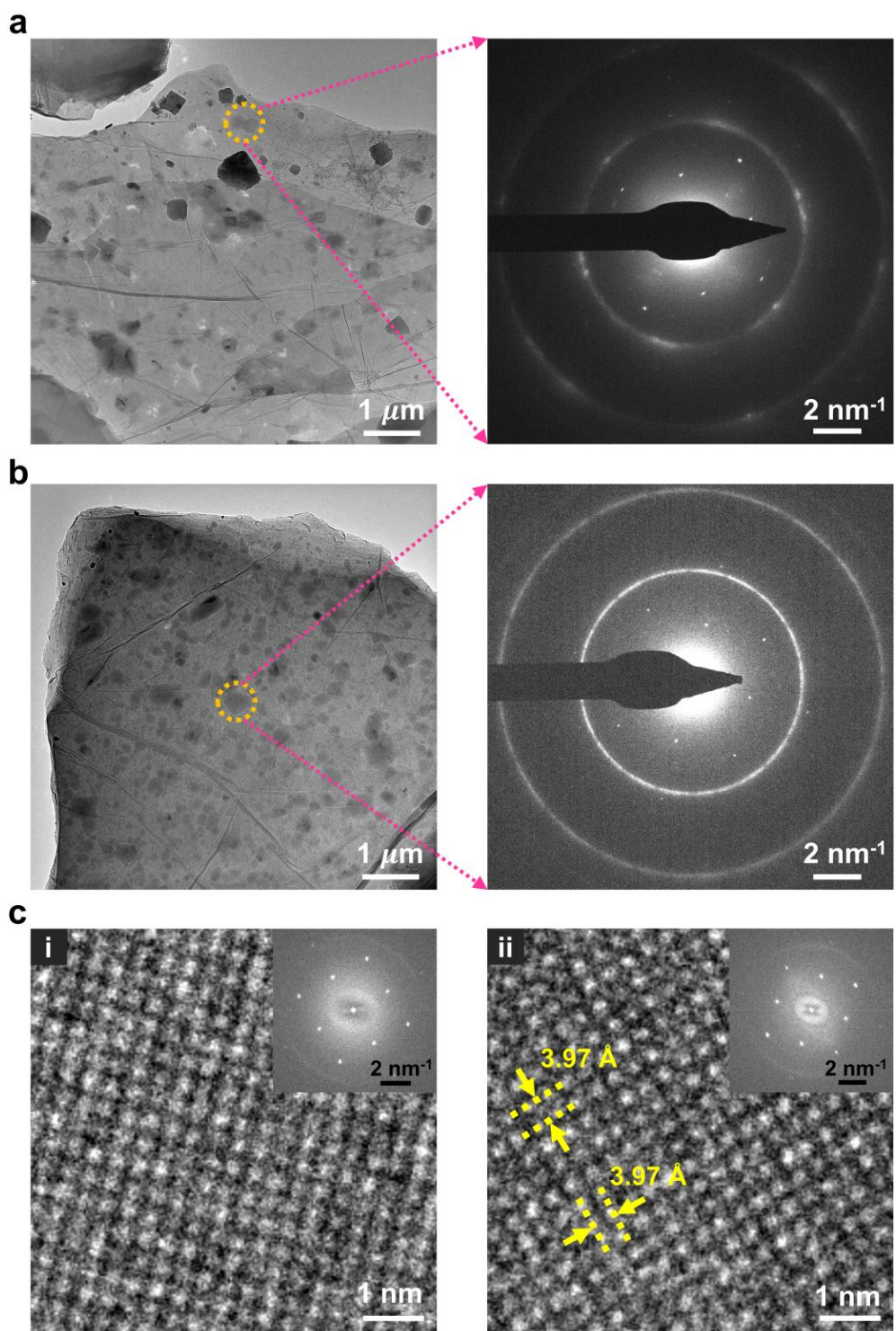

**Supplementary Fig. 1 | *In situ* characterization of Na–Cl crystals by cryo-EM. a** and **b** show the cryo-EM images of the different thin areas at the edge of Na–Cl@rGO membranes, the *in situ* distribution of Na–Cl crystals between rGO layers can be observed (left), and the diffraction patterns of the crystal domains are obtained in the electron diffraction mode (right). **c**, High-resolution atomic images of Na–Cl crystals obtained in the framed areas of **a** and **b** by cryo-EM.

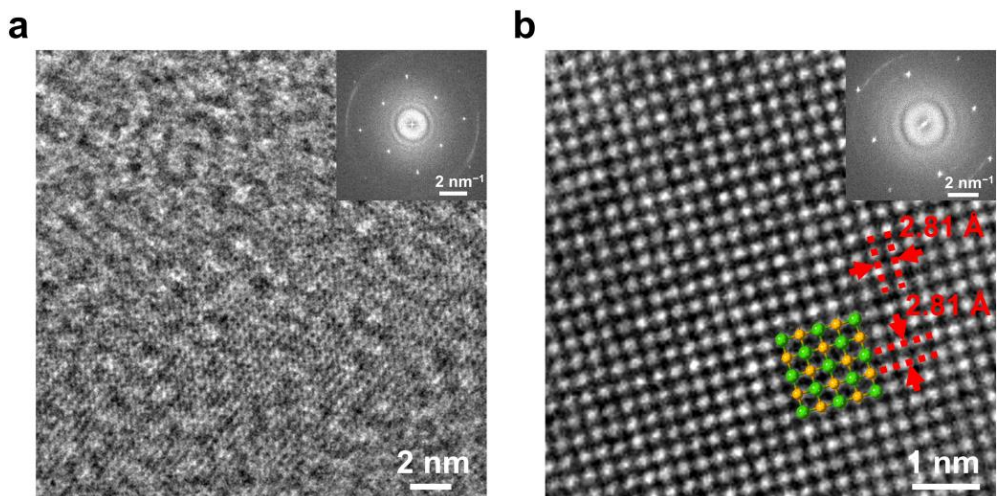

**Supplementary Fig. 2 | Conventional TEM images of special Na–Cl crystals and ordinary NaCl crystals taken on the dried Na–Cl@rGO membranes. a,** The special Na–Cl crystal observed at the edge of dried Na–Cl@rGO membranes without frozen process, the upper right illustration shows the FFT of the corresponding image, which shows the same square lattice as observed in Supplementary Fig. 1. **b,** The (200) crystal surface of ordinary NaCl crystals observed on the surface of dried Na–Cl@rGO membranes without frozen process, which are clearly distinct from the crystal structure that we observed under freezing conditions.

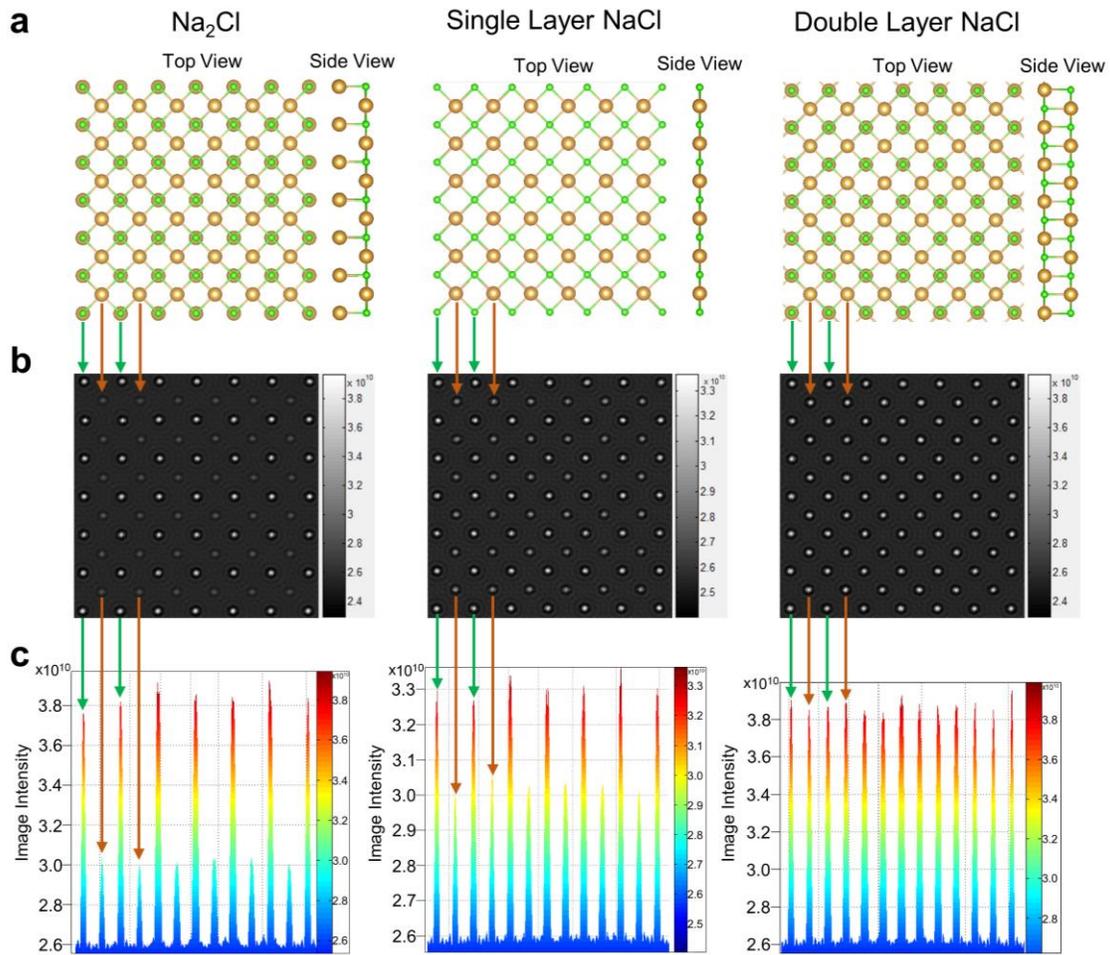

**Supplementary Fig. 3 | a**, Structural model of the recently predicted Na$_2$Cl crystal structure and the monolayer and double-deck structures of ordinary NaCl (200) crystal surface. **b,** TEM simulation images of the corresponding structure models in **a** obtained by QSTEM. **c,** The contrast of each atom position in the TEM simulation images of **b**.

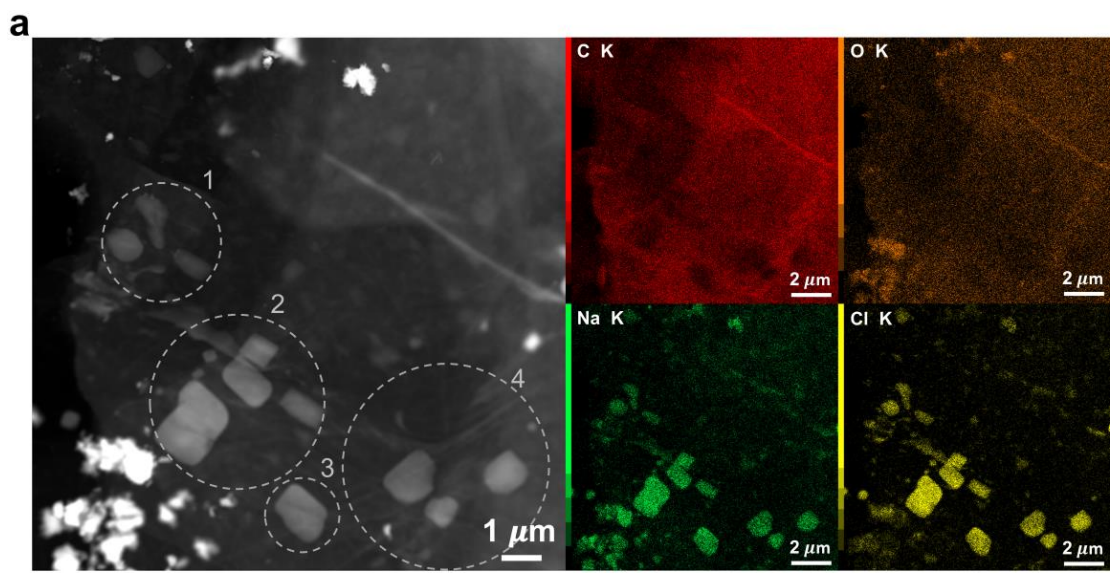

| Count | Na% | Cl% | Na:Cl |
|---|---|---|---|
| 1 | 3.71 | 3.16 | 1.17 |
| 2 | 6.94 | 6.61 | 1.05 |
| 3 | 7.46 | 6.56 | 1.14 |
| 4 | 2.48 | 1.77 | 1.40 |
| Whole area | 2.14 | 1.69 | 1.27 |

**Supplementary Fig. 4 | Elemental analysis of a region from dried rGO membranes after incubating in 3.0 M NaCl solution overnight. a**, HADDF-STEM image and the EDS mapping of this region and selected areas for quantitively elemental analysis. **b**, Table of the atomic ratio of Na to Cl measured in the selected areas of the dried Na–Cl@rGO membranes.

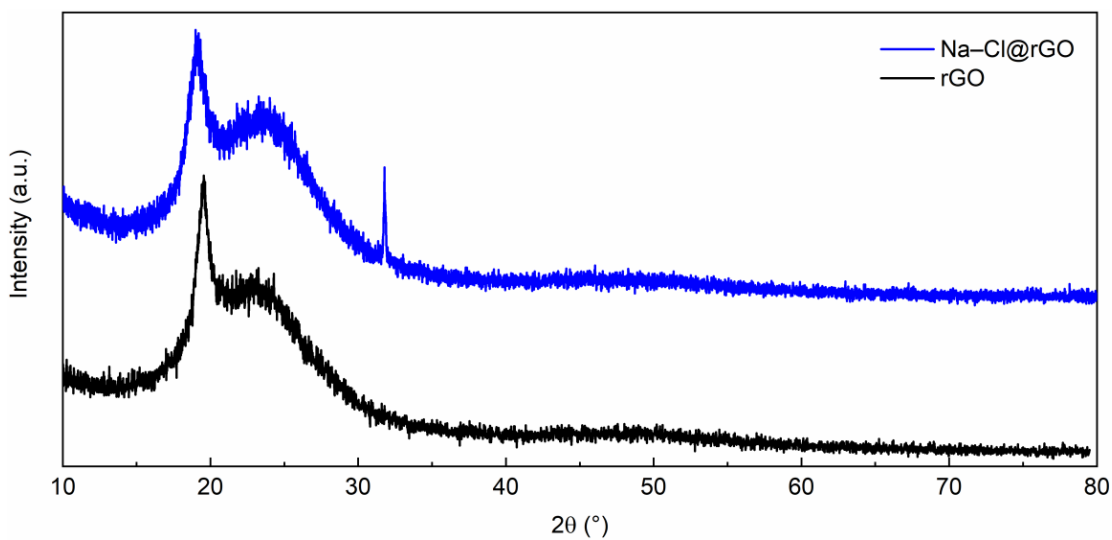

**Supplementary Fig. 5 | XRD patterns of pure rGO membranes and dried Na–Cl@rGO membranes.**

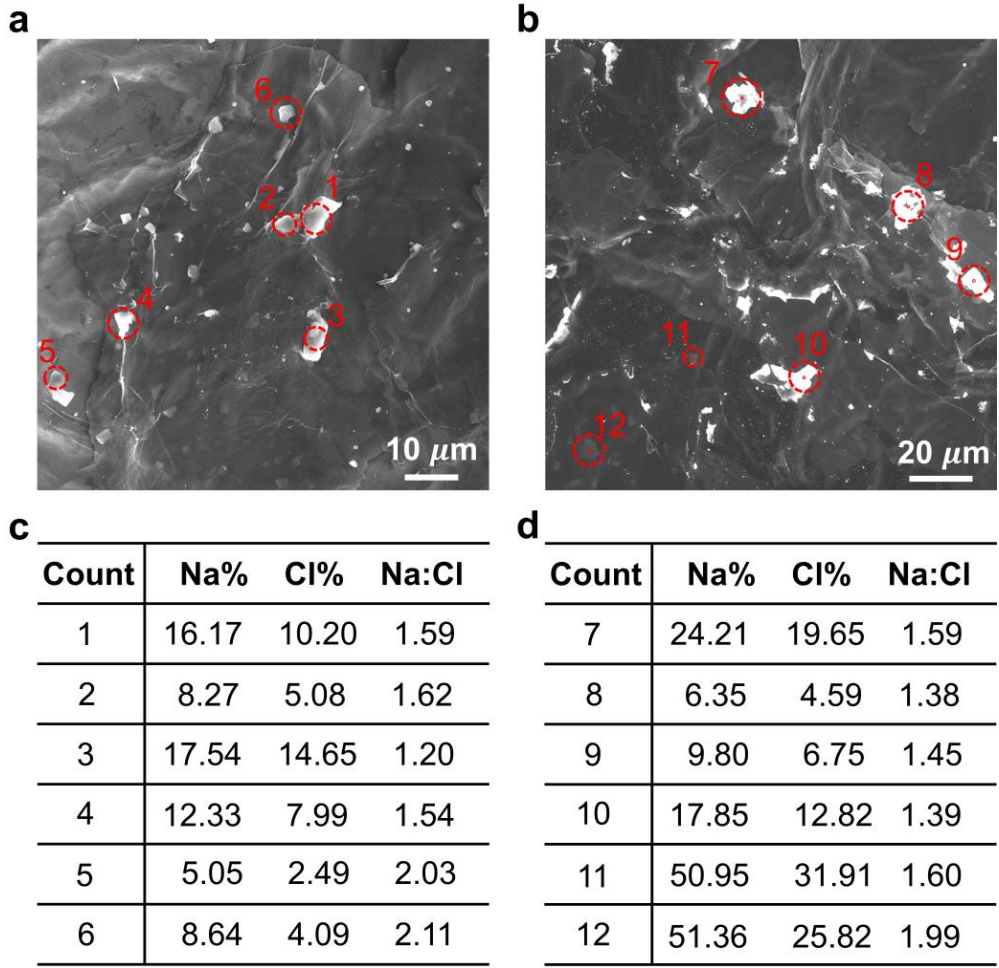

**Supplementary Fig. 6 | Elemental analysis by SEM of the torn dried Na–Cl@rGO membranes. a**, **b**, SEM topography image of the torn dried Na–Cl@rGO membranes. **c**, Table of the atomic ratio of Na to Cl measured in the selected areas of **a** by the "surface scan" mode. **d**, Table of the atomic ratio of Na to Cl measured in the selected areas of **b** by the "point scan" mode.

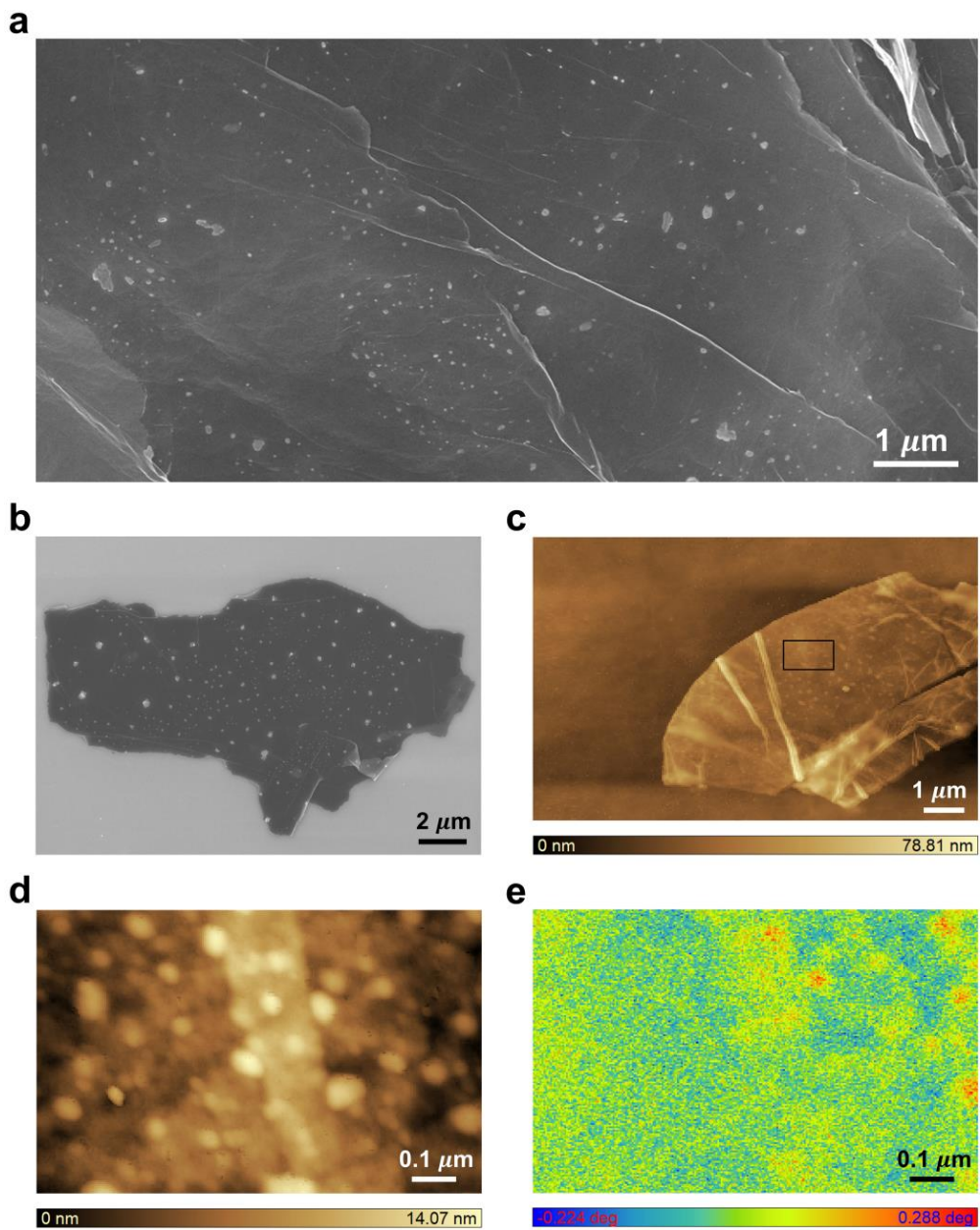

**Supplementary Fig. 7 | Sample preparation process for MFM-based magnetic characterization. a**, SEM topography image of the dried Na–Cl@rGO membranes after peeling off the membranes with tape using mechanical peeling method. **b**, SEM topography image of the dried Na–Cl@rGO membranes transferred to silicon wafer. **c**, AFM topography image of the dried Na–Cl@rGO membranes transferred to silicon wafer. **d**, AFM image of the selected area in **c**. **e**, MFM image obtained in the same area of **d** after raising the scanning probe by 50 nm.
8

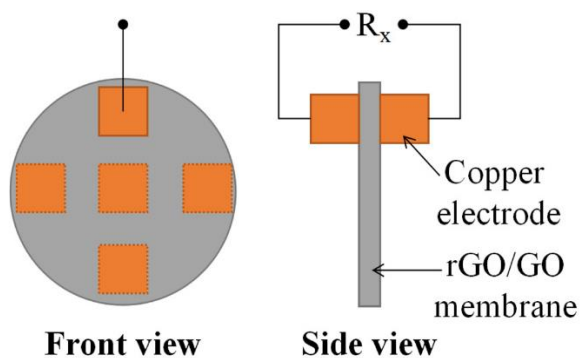

**Supplementary Fig. 8 | Schematic of the experimental setup for electrical resistivity measurement of the dried rGO or GO membranes.** Electrical resistivity of 5 positions measured by applying the multimeter with two copper electrodes connecting on the two side surfaces of the membranes.

**Supplementary Table 1 | Electrical resistivities of rGO and GO membranes**

| Membranes | 5 positions of electric resistivity (MΩ·cm) | | | | | Average (MΩ·cm) | Std. Dev. (MΩ·cm) |
|---|---|---|---|---|---|---|---|
| | (1) | (2) | (3) | (4) | (5) | | |
| GO-1 | $8.0\times10^3$ | $1.8\times10^4$ | $2.1\times10^4$ | $1.6\times10^4$ | $2.1\times10^4$ | | |
| GO-2 | $2.3\times10^4$ | $2.5\times10^4$ | $1.9\times10^4$ | $9.8\times10^3$ | $1.8\times10^4$ | $1.9\times10^4$ | $2.5\times10^4$ |
| GO-3 | $5.8\times10^3$ | $1.7\times10^4$ | $2.8\times10^4$ | $2.1\times10^4$ | $2.9\times10^4$ | | |
| Na–Cl@GO-1 | $5.6\times10^2$ | $9.7\times10^2$ | $4.8\times10^2$ | $3.6\times10^2$ | $3.7\times10^2$ | | |
| Na–Cl@GO-2 | $7.9\times10^2$ | $2.8\times10^2$ | $5.9\times10^2$ | $7.1\times10^2$ | $3.8\times10^2$ | $7.4\times10^2$ | $1.3\times10^3$ |
| Na–Cl@GO-3 | $1.4\times10^3$ | $1.1\times10^3$ | $1.1\times10^3$ | $9.5\times10^2$ | $1.1\times10^3$ | | |
| rGO-1 | $3.5\times10^{-2}$ | $3.2\times10^{-2}$ | $3.2\times10^{-2}$ | $4.0\times10^{-2}$ | $4.1\times10^{-2}$ | | |
| rGO-2 | $1.8\times10^{-2}$ | $2.6\times10^{-2}$ | $3.4\times10^{-2}$ | $4.0\times10^{-2}$ | $2.2\times10^{-2}$ | $3.3\times10^{-2}$ | $2.7\times10^{-2}$ |
| rGO-3 | $3.7\times10^{-2}$ | $3.5\times10^{-2}$ | $4.4\times10^{-2}$ | $2.7\times10^{-2}$ | $3.5\times10^{-2}$ | | |
| Na–Cl@rGO-1 | $1.4\times10^{-2}$ | $1.8\times10^{-2}$ | $2.0\times10^{-2}$ | $1.4\times10^{-2}$ | $1.5\times10^{-2}$ | | |
| Na–Cl@rGO-2 | $1.6\times10^{-2}$ | $1.3\times10^{-2}$ | $1.1\times10^{-2}$ | $1.5\times10^{-2}$ | $1.2\times10^{-2}$ | $1.4\times10^{-2}$ | $9.2\times10^{-3}$ |
| Na–Cl@rGO-3 | $1.3\times10^{-2}$ | $1.2\times10^{-2}$ | $1.2\times10^{-2}$ | $1.6\times10^{-2}$ | $1.4\times10^{-2}$ | | |